\documentclass[runningheads]{llncs}
\usepackage[T1]{fontenc}
\usepackage{amsfonts,amsmath,amssymb,bbm}
\usepackage[linesnumbered,ruled,vlined]{algorithm2e}
\usepackage{tikz}
\usetikzlibrary{tikzmark,calc}
\definecolor{pastelblue}{RGB}{179,205,227}   
\definecolor{pastelgreen}{RGB}{227, 238, 226} 
\definecolor{pastelpeach}{RGB}{253,222,200} 
\definecolor{pastelred}{RGB}{251,180,174} 
\usepackage{graphicx,enumitem}
\usepackage{caption}
\usepackage{subcaption}
\newcommand{\Kset}{\mathcal{K}}
\newcommand{\Kcomp}{\mathcal{K}_{\mathrm{comp}}}

\SetKwFor{ForEach}{for each}{do}{end}
\begin{document}
\title{StratFL: Countering Covert Model Poisoning Attacks with Strategic Federated Learning}
\titlerunning{StratFL}
%
\author{Manoj Twarakavi 
\and
Venkata Sriram Siddhardh Nadendla} 
\authorrunning{M. Twarakavi et al.}
%
\institute{Missouri University of Science and Technology, Rolla MO 65409, USA \\ 
\email{\{vthhw, nadendla\}@mst.edu}}
\maketitle              
\begin{abstract}
Advanced persistent threats (APTs) are a major concern in federated learning (FL) deployments in critical sectors such as healthcare, and finance. Traditional defense mechanisms fail to contain APTs due to their intermittent attacking behavior. This paper addresses a specific type of APT called \emph{covert model poisoning} (CMP) where the attacker intermittently injects poisoned model updates in a covert manner. The interaction between the FL server and a CMP attacker is modeled as an imperfect repeated game where both players cannot observe each other's actions. A novel adversarial search algorithm called \emph{StratFL} is designed alongside novel belief update algorithms at both the server and adversary. Experiments are designed using MNIST and CIFAR-10 datasets to evaluate and compare \emph{StratFL}'s performance against state-of-the-art methods. Empirical convergence of belief update algorithms is demonstrated across various security vignettes. At the same time, \emph{StratFL} is found to exhibit an improved accuracy of 10\% compared to state-of-the-art methods.

\end{abstract}
\section{Introduction}

Federated learning (FL) is a collaborative machine learning paradigm  in which distributed clients train models on their local data and transmit only the resulting model updates (to ensure data privacy) to a central server, which aggregates them into a shared global model \cite{fl}. Due to its improved learning performance and data privacy guaranties, FL has been adopted in many privacy-sensitive applications such as healthcare \cite{health_fl}, autonomous driving \cite{auto_fl}, and personal assistants (e.g. Google's GBoard \cite{gboard_fl} for text prediction, Apple's Siri \cite{siri_fl} for audio prediction). However, FL is still susceptible to \emph{advanced persistent threats} (APTs) where attackers exhibit patience and perform detrimental actions intermittently to create maximal damage to FL's learning performance. For example, the attacker could intermittently inject erroneous updates (a.k.a. \emph{poisoning}) across multiple training rounds (a.k.a \emph{training phase}) over a longer time-horizon. APTs have caused significant losses and have been under-investigated due to the temporal complexity of the attack and extensive resources available at the attacker. In 2025 alone, AI-driven APTs totaled to 16,200 incidents and costed \$5.72 million as per \cite{ai-cost}, with 55\% of these incidents targeted towards healthcare and finance sectors where FL is deployed extensively.  


Model poisoning is traditionally investigated as a one-shot threat [\cite{untargeted1}, \cite{untargeted2}, \cite{untargeted3},\cite{untargeted4}, \cite{untargeted5}, \cite{targeted1}, \cite{targeted2}, \cite{targeted3}, \cite{krum}, \cite{backgradient}] where the attacker reveals falsified model updates to the server by compromising a subset of clients through malware, man-in-the-middle (MITM) attack or other threat vectors. Furthermore, the past literature also investigated different types of model poisoning based on the underlying intent/motive behind the attack. For example, in \emph{untargeted model poisoning} [\cite{untargeted1}, \cite{untargeted2}, \cite{untargeted3}, \cite{untargeted4}, \cite{untargeted5}], the adversary's intent is to degrade the overall performance of the global model. In contrast, the adversary's intent behind \emph{targeted model poisoning} [\cite{targeted1}, \cite{targeted2}, \cite{targeted3}] is to misclassify a chosen set of inputs while leaving all other inputs intact. However, there is little/no attempt to investigate persistent versions of model poisoning where the attacker performs model poisoning intermittently to cause maximal impact on FL. This paper investigates a specific kind of APT called \emph{covert model poisoning} (CMP) where the attacker strategically injects poisoned model updates intermittently across time so that it remains undetected at the FL server.

In the past literature, various robust aggregation techniques [\cite{krum}, \cite{mean-median}, \cite{fltrust}, \cite{robustagg}, \cite{contra}, \cite{mudhog}] have been proposed to counter one-shot model poisoning at the FL server. These aggregation techniques can be classified into three broad categories: (i) \emph{robust statistics-based aggregation} to detect and suppress statistical outliers (e.g. Krum \cite{krum}, Trimmed-Mean \cite{mean-median}), (ii) \emph{selective aggregation} where deviations in client behavior over time are detected and removed from aggregation rules at the FL server (e.g. Auror \cite{auror}, FoolsGold \cite{cosine}), and (iii) \emph{benchmarked aggregation} where server compares the client updates against a trusted gradient (obtained using a benchmarked training set at the server) to screen-out misbehaving clients from FL aggregation (e.g. FLTrust \cite{fltrust}). While the latter two categories of robust aggregation rules can alleviate persistent model poisoning, they fail to mitigate CMP threats since the attacker remains hidden strategically for most time, and launches its attack only when the time is ripe. This paper develops a strategic aggregation rule at the FL server to identify and counter CMP.

The main contributions of this paper are three-fold. Firstly, this paper models the interaction between the FL server and the CMP attacker as an imperfect information game where both players are unaware of their opponent's strategies. This lack of information forces both players to adopt a novel belief updating algorithm to construct a conditional belief about the opponent's action respectively based on the history of observed losses computed using private validation datasets available at each player. Secondly, a novel adversarial search algorithm called \emph{StratFL} is proposed which computes the best responses using alternating projected gradient descent/ascent and the updated action beliefs at the FL server and attacker respectively. Thirdly, numerical experiments are designed based on standard image classification datasets to compare StratFL against a few state-of-the-art FL aggregation algorithms to investigate the performance in terms of accuracy change as well as the belief convergence regarding client types at the server.

The remainder of this paper is organized as follows: Section 2 establishes the threat model and the aggregation rule of the system. Section 3 defines the problem statement. Section 4 models the belief updating algorithms of the server and adversary. Section 5 describes the experimental setup. Section 6 details the numerical results, comparison, and their analysis. Section 7 concludes the paper.

\section{Threat Model}

\begin{figure}[!t]
\centering
\includegraphics[width=\textwidth]{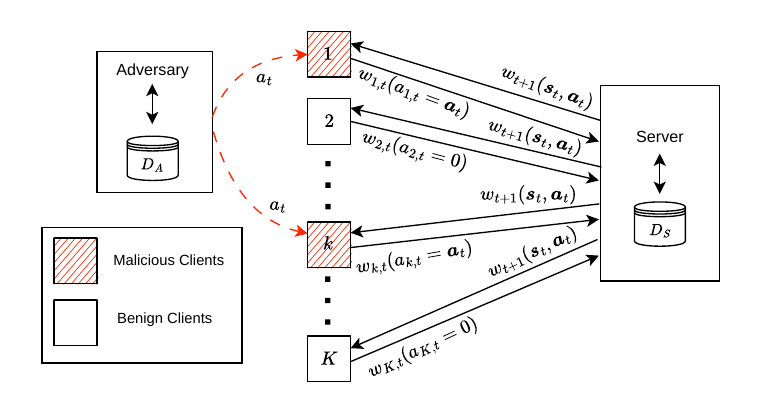}
\caption{Strategic Federated Learning and Covert Model Poisoning}
\label{strategicfl}
\end{figure}

Consider an FL system as illustrated in Figure \eqref{strategicfl}, which consists of a client set $\mathcal{K} = \{ 1, \cdots, K \}$ and a central server. Assume that the FL system is designed to learn a classifier that characterizes the relationship between the input space $\mathbb{R}^M$ and the output space $\mathcal{Y} = \{1, \cdots, N\}$. Let the classifier in the FL system be a neural network (NN) $f$ that returns a probabilistic output 
\begin{equation}
\hat{y} = f (\boldsymbol{x} | w ) \quad \in \ \mathcal{S}_N
\label{eq:1}
\end{equation}
in one-hot form where $\mathcal{S}_N$ is the probability simplex defined in the discrete output space $\mathcal{Y}$, and $w \in \mathbb{R}^L$ are the parameters of the NN model. Given any dataset $D$ and an instantaneous loss function $\ell: \mathcal{Y} \times \mathcal{S}_N \rightarrow \mathbb{R}_+$, the average loss of a model $f$ with parameters $w$ can be evaluated as
\begin{equation}
\mathcal{L}(w,D) = \displaystyle \frac{1}{|D|} \sum_{(\boldsymbol{x}, y) \in D} \ell \Big( y, f( \boldsymbol{x} | w ) \Big).
\label{Eqn: average loss given dataset}
\end{equation}
where $y$ is the true label corresponding to the input $\boldsymbol{x}$ in $D$, and $|D|$ represents the total number of samples in $D$.

FL is an iterative learning framework where clients collect new data privately to train their local models and share the model parameters to the server. The server aggregates these local models to obtain a global model $w_t$ at round $t$. Assume that the client $k$ possesses a private dataset $D_{k,t} = \{ (\boldsymbol{x}_{k, 1:tN}, y_{k,1:tN}) \}$ at round $t$. In other words, the $k^{th}$ client elicits $N$ private data samples 
$$(\boldsymbol{x}_{k,(t-1)N + 1}, y_{k,(t-1)N + 1}), \cdots, (\boldsymbol{x}_{k,tN}, y_{k,tN})$$ 
at the beginning of round $t$ and appends it to its previous dataset $D_{k,t-1}$ to obtain $D_{k,t}$. The $k^{th}$ client computes the loss gradient 
\begin{equation}
g_{k,t} = \displaystyle \nabla_{w} \mathcal{L}(w,D_{k,t}) \Big|_{w = w_{t-1}}
\label{eq:client loss gradient}
\end{equation}
using the global model $w_{t-1}$, where the loss $\mathcal{L}(w, D_{k,t})$ can be computed using Equation \eqref{Eqn: average loss given dataset}. 

Assume that the CMP attacker controls $\alpha$ fraction of clients in the FL system. Let the $k^{th}$ client type be denoted as
\begin{equation}
\gamma_{k} = 
\begin{cases}
1, & \text{if the $k^{th}$ client is controlled by the CMP attacker},
\\[0.5ex]
0, & \text{otherwise}.
\end{cases}
\end{equation}
If the $k^{th}$ client is compromised (i.e. $\gamma_k = 1$), the attacker chooses a stochastic action $a_{k,t} \in [0,1]$ at round $t$ to modify the loss gradient $g_{k,t}$ in Equation \eqref{eq:client loss gradient} as 
\begin{equation}
\begin{array}{lcl}
\tilde{g}_{k,t}(a_{k,t}) & = & \displaystyle  a_{k,t} \cdot \hat{g}_{k,t} +  (1 - a_{k,t}) \cdot g_{k,t}
\\[2ex]
& = & (1 - 2a_{k,t}) \cdot g_{k,t}
\end{array}
\label{eq:gradient-attack}
\end{equation}
where $\hat{g}_{k,t} = - g_{k,t}$ is the negative loss gradient used to flip the $k^{th}$ client's motive towards loss maximization. Note that such an attack model can be generalized to explain the behavior of all clients, since a benign client would always choose $a_{k,t} = 0$.

Given the loss gradient $\tilde{g}_{k,t}(a_{k,t})$, assume that the $k^{th}$ client updates its model using gradient descent algorithm as
\begin{equation}
w_{k,t}(a_{k,t}) = w_{t-1} - \eta \cdot \tilde{g}_{k,t}(a_{k,t}).
\label{Eqn: Compromised node - Model update eqn}
\end{equation}
where $\eta$ is the learning rate\footnote{Note that the model can be easily extended to have heterogeneous learning rates at the clients which could be true in general. However, this case is not within the scope of this study and is therefore excluded from this paper.} adopted by the clients. Subsequently, the clients share their updated local models to the server for aggregation.

Being unaware of the client types (benign/compromised), the server fuses these local models $w_{1,t}, \cdots, w_{K,t}$ using a \emph{selective aggregation rule} at round $t$ as
\begin{equation}
w_t(\boldsymbol{s}_t, \boldsymbol{a}_t) = \displaystyle \sum_{k = 1}^{K} s_{k,t} \cdot w_{k, t}(a_{k,t})
\label{Eqn: Server - Selective Aggregation Rule}
\end{equation}
where $s_{k,t} \in [0,1]$ represents the server's normalized selection weight assigned to the $k^{th}$ client's model. In other words, $\displaystyle \sum_{k = 1}^K s_{k,t} = 1$ for all $t$.

\section{Problem Statement}

Note that the global model $w_{t+1}$ at the server is a function of both the server decision tuple $\boldsymbol{s}_t = \{ s_{1,t}, \cdots, s_{K,t} \}$ and the CMP attacker's decision tuple $\boldsymbol{a}_t = \{ a_{k,t} \}_{k \in \mathcal{K}_{comp}}$. In most practical applications, the decisions made by the server and the CMP attacker are private and unobservable to the other agent. Therefore, it is natural to model the interaction between server and the CMP attacker as a imperfect repeated game, where both the agents construct beliefs about each other's actions and update them over time. Let 
$$\pi_t(\boldsymbol{a} | w_{k,1:t}) = \mathbb{P}(\boldsymbol{a}_t = \boldsymbol{a} \ | \ w_{1,1:t}, \cdots, w_{K,1:t})$$
denote the server's conditional belief that the attacker's decision sequence is $\boldsymbol{a}_t = \boldsymbol{a}$ given the past model parameters $w_{1,1:t}, \cdots, w_{K,1:t}$ at the server. Similarly, let 
$$\phi_t(\boldsymbol{s} | w_{1:t})  = \mathbb{P}(\boldsymbol{s}_t = \boldsymbol{s} | w_{1:t})$$ 
denote the attacker's conditional belief that the server's choice is $\boldsymbol{s}_t = \boldsymbol{s}$ given the sequence of shared global models $w_{1:t}$ over $t$ rounds. 

Assuming that the server possesses a private dataset $D_S$, the server evaluates its average loss at round $t$ as
\begin{equation}
\mathcal{L}_S(\boldsymbol{s}_t |  \pi_t) = \displaystyle \mathbb{E}_{\pi_t} \left[ \mathcal{L} \Big( w_t(\boldsymbol{s}_t, \boldsymbol{a}_t), {D}_{S} \Big) \right].
\label{Eqn: Server Objective}
\end{equation}
The main goal at the server is to find an appropriate $\boldsymbol{s}_t$ that minimizes the average loss $\mathcal{L}_S(\boldsymbol{s}_t | \pi_t)$. 

On the contrary, the attacker holds its own private validation dataset $D_A$ to compute its average loss
\begin{equation}
\mathcal{L}_A(\boldsymbol{a}_t |  \phi_t) = \displaystyle \mathbb{E}_{\phi_t} \left[ \mathcal{L} \Big( w_{t}(\boldsymbol{s}_{t}, \boldsymbol{a}_{t}), {D}_{A} \Big) \right].
\label{Eqn: Attacker loss}
\end{equation}
While the attacker wishes to maximize $\mathcal{L}_A(\boldsymbol{a}_t |  \phi_t)$ at round $t$, it also wishes to stay hidden as long as it can to launch a persistent threat. The server infers the type of $k^{th}$ type indirectly through its choice of $s_{k,t}$. For example, the server only chooses a small value for $s_{k,t}$ (close to zero) if it believes that the $k^{th}$ client is compromised, i.e. $\gamma_k = 1$. On the contrary, the server chooses $s_{k,t} = 1$ if it trusts the the $k^{th}$ client is benign. In other words, the potential of detecting the presence of a CMP attacker is quantified by $\displaystyle \sum_{k \in \mathcal{K}_{comp}} (1 - s_{k,t} )$. Therefore, the attacker's goal can be summarized as a tradeoff between maximizing $\mathcal{L}_A(\boldsymbol{a}_t |  \phi_t)$ and minimizing $\displaystyle \mathbb{E}_{\phi_t} \left[ \sum_{k \in \mathcal{K}_{comp}} (1 - s_{k,t} ) \right]$. 

In summary, the strategic interaction between the server and the CMP attacker can be modeled as the following imperfect repeated game:
\begin{equation}
\begin{array}{lcl}
\boldsymbol{s}_t^* & = & \displaystyle \min_{\boldsymbol{s}_t \in [0,1]^K} \ \mathcal{L}_{S}(\boldsymbol{s}_t |  \pi_t)   
\\[1.5ex]
\boldsymbol{a}_t^* & = & \displaystyle  \max_{\boldsymbol{a}_t \in [0,1]^{|\mathcal{K}_{comp}|}} \  \mathcal{L}_{A}(\boldsymbol{a}_t |  \phi_t) - \lambda \cdot \mathbb{E}_{\phi_t} \left[ \sum_{k \in \mathcal{K}_{comp}} (1 - s_{k,t} ) \right] 
\end{array}
\tag{P1}
\label{Prob: game}
\end{equation}
where $\lambda$ quantifies the CMP attacker's tradeoff between loss maximization and its desire to stay covert.

\section{Strategic Federated Learning}

Prior to solving Problem \ref{Prob: game}, it is essential to clarify how both agents update their beliefs over time. Sections \ref{sec:server-beliefupdate} and \ref{sec: Attacker Belief Update} present novel belief updating algorithms at the server and the attacker respectively. Then, Section \ref{sec: Adversarial Search} presents an adversarial search based on the novel belief updating algorithms at the server and the attacker.

\subsection{Server's Belief Updating}
\label{sec:server-beliefupdate}

Let $\ell_{t-1} = \mathcal{L} ( w_{t-1}, D_S )$ denote the average loss at the server at the beginning of round $t$. Upon receiving all the clients' local models $w_{k,t}$ for all $k = 1, \cdots, K$, the server computes the $k^{th}$ client's loss as $\ell_{k,t} = \mathcal{L} ( w_{k, t}, D_S)$. After aggregating the local client models into a global model $w_t$, the new average loss at the server is given by $\ell_{t} = \mathcal{L} ( w_{t}, {D}_{S} )$. 

The spirit behind the FL design is that the model performance at the server can be improved through the fusion of several weak models obtained from the clients. In other words, it is natural to expect that the client model $w_{k,t}$ performs better than $w_{t-1}$ due to local training on $D_{k,t}$, while the global model $w_t$ performs better than $w_{k,t}$ due to model aggregation. In summary, there should exist a small loss tolerance $\epsilon > 0$ so that the server expects the following inequality to hold true:
\begin{equation}
\ell_{t} < \ell_{k,t} < \ell_{t-1} (1-\epsilon).
\label{eq:inequality-server}
\end{equation}

Taking inspiration from the above inequality, let the loss deviation at client $k$ in round $t$ be defined as $d_{k,t} = \mathbbm{1} \Big( \ell_{t-1} (1-\epsilon) - \ell_{k,t} \Big)$, where $\mathbbm{1}(e)$ denotes an indicator function which returns 1 if the expression $e$ is positive, and 0 otherwise. If $d_{k,t} = 0$, the $k^{th}$ client's model $w_{k,t}$ has worse performance than the global model $w_{t-1}$, which can only happen if the $k^{th}$ client is compromised [refer to Equation \eqref{eq:gradient-attack}]. The server can eventually count such deviations at each client over multiple rounds to update its beliefs. 

Since the indicator function in the deviation $d_{k,t}$ is not differentiable, the deviations are unusable in their original form. Therefore, we consider a continuous surrogate to the deviation as defined below.

\begin{definition}[Server's Surrogate Loss Deviation]
The surrogate loss deviation is a sigmoidal approximation of the loss deviation, which can be formally defined as
\begin{equation}
c_{k,t} = \sigma \Big( \ell_{t-1} (1-\epsilon) - \ell_{k,t} \Big)
\label{Eqn: Surrogate Loss Deviation}
\end{equation}

\label{Defn: Surrogate Loss Deviation}
\end{definition}

Given the past $T$ surrogate loss deviations $c_{k,t-T+1}, \cdots, c_{k,t}$, the server can estimate the marginal probability that the $k^{th}$ client is compromised (and therefore choose $a_{k,t} = 1$) as
\begin{equation}
p_{k,t} = \displaystyle \frac{1}{T} \sum_{\tau = 1}^T c_{k,t-T+\tau}.
\label{Eqn: belief about client-k type}
\end{equation}

Note that Equation \eqref{Eqn: belief about client-k type} can also be computed using a linear recursion as
\begin{equation}
p_{k,t} = \displaystyle p_{k,t-1} + \Delta_{k,t},
\label{Eqn: belief about client-k type - recursive}
\end{equation}
where $\Delta_{k,t} = \displaystyle \frac{1}{T} \Big( c_{k,t} - c_{k,t-T} \Big)$ is the non-recursive term.

Given the marginal attack probabilities $p_{1,t}, \cdots, p_{K,t}$ at all the clients in round $t$, the server's conditional joint belief regarding the attacker's decision sequence $\boldsymbol{a}_t = \{ a_{1,t}, \cdots, a_{K,t} \}$ is given by
\begin{equation}
\pi_t(\boldsymbol{a}_t) = \displaystyle \prod_{k=1}^{K} p_{k,t}^{a_{k,t}} (1 - p_{k,t})^{1 - a_{k,t}}
\label{Eqn: pi_t - noniterative}
\end{equation}

The server's belief updating algorithm is summarized in Algorithm \ref{alg:server}.

\begin{algorithm}[!t]
\caption{Server Belief Update} 
\label{alg:server}
\DontPrintSemicolon
\SetKwFunction{OptSel}{BeliefUpdate}
\SetKwFunction{Server}{Server}
\SetKwFunction{Client}{Client}
\SetKwFunction{VL}{ComputeValidationLoss}
\SetKwFunction{LL}{ComputeLocalLoss}
\SetKwFunction{GL}{ComputeGlobalLoss}
\SetKwFunction{Belief}{ComputeBelief}
\SetKwFunction{OptSel}{ServerBeliefUpdate}
\SetKwProg{Fn}{}{:}{end}
\Fn{\OptSel{$w_{t-1}, \{w_{k,t}\}_{k=1}^{K}$}}{
  $l_{t-1}\gets \Server.\VL (w_{t-1},D_S)$\; 
  \BlankLine
  \tikzmark{stop}%
  \ForEach{$k\in\Kset$}{
    $l_{k,t}\gets\Server.\LL(w_{k,t},D_S)$ \;
    $c_{k,t}\gets$ Compute Using Equation \eqref{Eqn: Surrogate Loss Deviation} \;
    $p_{k,t}\gets$ Compute using Equation \eqref{Eqn: belief about client-k type}\;
    $\pi_t\gets$  Compute using Equation \eqref{Eqn: pi_t - noniterative}\;
  }
  \tikzmark{sbot}%
  \KwRet $\pi_{t}(\boldsymbol{a}_{t})$\;
}
\end{algorithm}
\begin{tikzpicture}[remember picture, overlay]
  \coordinate (BoxNW) at ([xshift=-0.5em, yshift=2.5ex]pic cs:stop);
  \coordinate (BoxSE) at ([xshift=0.6\linewidth, yshift=2.2ex]pic cs:sbot);
  \begin{scope}[blend mode=multiply]
    \fill[rounded corners, pastelblue] (BoxNW) rectangle (BoxSE);
  \end{scope}
  \node[anchor=west, xshift=0.8em, text width=0.15\linewidth,
        align=left, font=\scriptsize]
    at (BoxSE |- {$(BoxNW)!0.5!(BoxSE)$})
    {};
\end{tikzpicture}

\subsection{Attacker's Belief Updating \label{sec: Attacker Belief Update}}

Let $\tilde{\ell}_t = \mathcal{L} \Big( w_{t}, {D}_{A} \Big)$ denote the server's average loss at the CMP attacker computed using the attacker's private validation dataset $D_A$. Similarly, at the $k^{th}$ client (assuming that it is compromised), the CMP attacker can evaluate the client model's loss as $\tilde{\ell}_{k,t} = \mathcal{L} \Big( w_{k,t}, {D}_{A} \Big)$ upon updating its local model. Similar to the inequality used by the server in Equation \eqref{eq:inequality-server}, there should exist a small loss tolerance $\epsilon > 0$ so that the attacker expects the following inequality to be true:
\begin{equation}
\tilde{\ell}_{t}(1-\epsilon) < \tilde{\ell}_{k,t} < \tilde{\ell}_{t-1}.
\label{eq:inequality-attacker}
\end{equation}

Note that the attacker wishes to evaluate the belief about the server's selection strategy. The loss deviation $\tilde{d}_{k,t} = \mathbbm{1}(\tilde{\ell}_t(1-\epsilon) - \tilde{\ell}_{k,t})$ captures the server's decision $s_{k,t}$ such that $\tilde{d}_{k,t}$ decreases to zero as $s_{k,t}$ decreases to zero. Since the deviation $\tilde{d}_{k,t}$ is non-differentiable, the surrogate loss deviation is adopted by the attacker as defined below.
\begin{definition}[Attacker's Surrogate Loss Deviation]
The surrogate loss deviation at the attacker is a sigmoidal approximation of the loss deviation, which can be formally defined as
\begin{equation}
\tilde{c}_{k,t} = \sigma \Big( \tilde{\ell}_{t} (1-\epsilon) - \tilde{\ell}_{k,t} \Big)
\label{Eqn: Attacker's Surrogate Loss Deviation}
\end{equation}
\end{definition}

Based on the past $T$ surrogate loss deviations $\tilde{c}_{k,t-T+1}, \cdots, \tilde{c}_{k,t}$, the attacker estimates the marginal probability that the $k^{th}$ client is selected by the server (and therefore choose $s_{k,t} = 1$) as
\begin{equation}
q_{k,t} = \displaystyle \frac{1}{T} \sum_{\tau = 1}^T \tilde{c}_{k,t-T+\tau}.
\label{Eqn: belief about client-k selection}
\end{equation}

Note that Equation \eqref{Eqn: belief about client-k selection} can also be computed using a linear recursion as
\begin{equation}
q_{k,t} = \displaystyle q_{k,t-1} + \tilde{\Delta}_{k,t},
\label{Eqn: belief about client-k selection - recursive}
\end{equation}
where $\tilde{\Delta}_{k,t} = \displaystyle \frac{1}{T} \Big( \tilde{c}_{k,t} - \tilde{c}_{k,t-T} \Big)$ is the non-recursive term.

Given the marginal attack probabilities $q_{1,t}, \cdots, q_{K,t}$ in round $t$, the attacker's conditional joint belief regarding the server's decision sequence $\boldsymbol{s}_t = \{ s_{1,t}, \cdots, s_{K,t} \}$ is given by
\begin{equation}
\phi_t(\boldsymbol{s}_t) = \displaystyle \prod_{k \in \mathcal{K}_{comp}} q_{k,t}^{s_{k,t}} (1 - q_{k,t})^{1 - s_{k,t}}.
\label{Eqn: phi_t - noniterative}
\end{equation}

This belief updating at the attacker is summarized in Algorithm \ref{alg:attacker}.

\begin{algorithm}[!t]
\caption{Adversary Belief Update} 
\label{alg:attacker}
\DontPrintSemicolon
\SetKwFunction{OptSel}{BeliefUpdate}
\SetKwFunction{Server}{Server}
\SetKwFunction{Client}{Client}
\SetKwFunction{Adv}{Adversary}
\SetKwFunction{VL}{ComputeValidationLoss}
\SetKwFunction{LL}{ComputeLocalLoss}
\SetKwFunction{GL}{ComputeGlobalLoss}
\SetKwFunction{Belief}{ComputeBelief}
\SetKwFunction{Poison}{PoisonModel}
\SetKwFunction{OptSel}{AdversaryBeliefUpdate}
\SetKwProg{Fn}{}{:}{end}
\Fn{\OptSel{$w_{t}, \{w_{k,t}\}_{k}^{\mathcal{K}_{comp}}$}}{
  $\tilde{\ell}_t\gets \Adv.\VL (w_t,D_A)$\;
  \BlankLine
  \tikzmark{atop}%
  \ForEach{$k\in\Kcomp$}{
    $\tilde{\ell}_{k,t}\gets \Adv.\LL(w_{k,t}, D_A) $\; 
    $\tilde{c}_{k,t}\gets$ Compute Using Equation \eqref{Eqn: Attacker's Surrogate Loss Deviation} \; 
    $\tilde{q}_{k,t}\gets$ Compute Using Equation \eqref{Eqn: belief about client-k selection} \; 
    $\phi_t\gets$ Compute Using Equation \eqref{Eqn: phi_t - noniterative}\;
    }
    \tikzmark{abot}%
  \KwRet $\phi_{t}(\boldsymbol{s}_{t})$
}

\begin{tikzpicture}[remember picture, overlay]
  \coordinate (BoxNW) at ([xshift=-0.5em, yshift=2.5ex]pic cs:atop);
  \coordinate (BoxSE) at ([xshift=0.62\linewidth, yshift=2.2ex]pic cs:abot);
  \begin{scope}[blend mode=multiply]
    \fill[rounded corners, pastelred] (BoxNW) rectangle (BoxSE);
  \end{scope}
  \node[anchor=west, xshift=0.8em, text width=0.1\linewidth,
        align=left, font=\scriptsize]
    at (BoxSE |- {$(BoxNW)!0.5!(BoxSE)$})
    {};
\end{tikzpicture}
\end{algorithm}

\subsection{StratFL: A Projected Gradient-based Adversarial Search \label{sec: Adversarial Search}}

\begin{algorithm}[!t]
\caption{Strategic Federated Learning (\textsc{StratFL})}
\label{alg:strategicfl}
\DontPrintSemicolon
\SetKwData{Server}{Server}
\SetKwData{Client}{Client}
\SetKwData{Adv}{Adversary}
\SetKwFunction{Init}{InitializeModelParams}
\SetKwFunction{Broadcast}{Broadcast}
\SetKwFunction{GetParams}{GetModelParams}
\SetKwFunction{OptAttack}{BeliefUpdate}
\SetKwFunction{OptSel}{BeliefUpdate}
\BlankLine
$w_{0} \gets$ \Server.\Init{$s_{k,0}\!\gets\!1,\;\forall k\in\Kset$}\;
\BlankLine
$w_{0} \gets$ \Adv.\Init{$a_{k,0}\!\gets\!0.5,\;\forall k\in\Kcomp$}\;
\BlankLine
\For{$t\gets 1$ \KwTo $T-1$}{
  \BlankLine
  \tikzmark{top}%
  \ForEach{$k\in\Kset$ \textnormal{\textbf{in parallel}}}{
    \uIf{$k\notin\Kcomp$}{
      $w_{k,t+1}\gets$ \Client.\GetParams{$w_t,D_{k,t}$} 
    }
    \Else{
      $\widetilde w_{k,t+1}\gets$ \Adv.\GetParams{$w_t, D_{k,t}, a_{k,t}$} 
    }
  }
  \tikzmark{bot}%
  \BlankLine
  \tikzmark{top2}%
  $a_{t+1}\gets$ \Adv.\OptAttack{$w_{t}, \{w_{k,t}\}_{k \in \mathcal{K}_{comp}}$} \tcp*{Alg.~\ref{alg:attacker}}
  $s_{t+1}\gets$ \Server.\OptSel{$w_{t-1}, \{w_{k,t}\}_{k \in \mathcal{K}}$} \tcp*{Alg.~\ref{alg:server}}
\tikzmark{bot2}%
  $w_{(t+1)}(\boldsymbol{s}_t, \boldsymbol{a}_t)\gets$ $\displaystyle \sum_{k = 1}^{K} \bigg( \frac{s_{k,t}}{\sum_{i=1}^{K} s_{i,t}} \bigg) \cdot w_{k, t+1}(a_{k,t})$\;
}
\KwRet $w_{(t+1)}(\boldsymbol{s}_t, \boldsymbol{a}_t)$
\begin{tikzpicture}[remember picture, overlay]
  \coordinate (BoxNW) at ([xshift=-0.75em, yshift=2.2ex]pic cs:top);
  \coordinate (BoxSE) at ([xshift=0.75\linewidth, yshift=1.2ex]pic cs:bot);
  \draw[thick, rounded corners] (BoxNW) rectangle (BoxSE);
  \node[anchor=west, xshift=0.8em, text width=0.15\textwidth,
        align=left, font=\tiny]
    at (BoxSE |- {$(BoxNW)!0.5!(BoxSE)$})
    {Local training through the shared \GetParams{}.};
\coordinate (BoxNW) at ([xshift=-0.75em, yshift=2.4ex]pic cs:top2);
 \coordinate (BoxSE) at ([xshift=0.9\textwidth, yshift=4.5ex]pic cs:bot2);
 \begin{scope}[blend mode=multiply]
    \fill[rounded corners, pastelgreen] (BoxNW) rectangle (BoxSE);
  \end{scope}
\end{tikzpicture}
\end{algorithm}

Given the belief updating algorithms at both the server and the attacker in Algorithms \ref{alg:server} and \ref{alg:attacker} respectively, the agents can now evaluate their respective objective functions in Problem \ref{Prob: game}. Therefore, this section focuses on the development of adversarial search algorithms at both the server and the attacker to solve Problem \ref{Prob: game}. Note that these adversarial search algorithms iterate alternatively between the attacker (leader) and the server (follower) since the attacker launches the attack before the server chooses its selection strategy.

\subsubsection{CMP Attacker's Best Response:}
In alignment with the attacker's goal to maximize its objective as stated in Problem \ref{Prob: game}, the attacker's best response $\boldsymbol{a}_t^*$ can be iteratively computed using the projected gradient ascent algorithm, as shown below:
\begin{equation}
\boldsymbol{a}_{t+1} = \mathbb{P}_{[0,1]^{\mathcal{K}_{comp}}} \Bigg[ \boldsymbol{a}_t + \eta_a \cdot \nabla_{\boldsymbol{a}_{t}} \left( \mathbb{E}_{\phi} \Big( \mathcal{L} ( w_{t+1}(\boldsymbol{s}_t, \boldsymbol{a}_t), {D}_{A} \Big) - \lambda \sum_{k \in \mathcal{K}_{comp}} (1 - s_{k,t} ) \right) \Bigg]
\label{eq:attacker-strategy-update}
\end{equation}

The challenge with computing the above projected gradient ascent step lies in the computation of the gradient $\nabla_{\boldsymbol{a}_t} s_{k,t}$. This is because $s_{k,t}$ is a best-response function for which gradient computation is cumbersome. Instead, we propose the replacement of $s_{k,t}$ with its surrogate $q_{k,t}$ which is computed in Equation \eqref{Eqn: belief about client-k selection}. In other words, the approximate best response at the attacker is computed as
\begin{equation}
\boldsymbol{a}_{t+1} \approx \mathbb{P}_{[0,1]^{\mathcal{K}_{comp}}} \Bigg[ \boldsymbol{a}_t + \eta_a \cdot \nabla_{\boldsymbol{a}_{t}} \left( \mathbb{E}_{\phi} \Big( \mathcal{L} ( w_{t+1}(\boldsymbol{s}_t, \boldsymbol{a}_t), {D}_{A} \Big) - \lambda \sum_{k \in \mathcal{K}_{comp}} (1 - q_{k,t} ) \right) \Bigg]
\label{Eqn: attacker best response approximate}
\end{equation}
where the gradient of $q_{k,t}$ can be computed using standard automatic differentiation libraries (e.g. Autograd in PyTorch).

\subsubsection{Server's Best Response:} The projected gradient descent at the server is straightforward and can be iteratively computed as follows.
\begin{equation}
\boldsymbol{s}_{t+1} = \mathbb{P}_{[0,1]^{K}} \Bigg[ \boldsymbol{s}_t - \eta_s \cdot \nabla_{\boldsymbol{s}_{t}} \mathcal{L}_S \Bigg],
\label{Eqn: server best response}
\end{equation}

Both the best responses at the server and attacker are summarized together into one adversarial search algorithm called \emph{StratFL}, which is presented in Algorithm \ref{alg:strategicfl}.


\section{Validation Methodology}
\label{sec: validation methodology}

The performance of \emph{StratFL} relies on all of its three algorithmic components: (i) server's belief updating, (ii) attacker's belief updating, and (iii) projected gradient-based adversarial search. In order to validate \emph{StratFL}, experiments are designed using well-known image classification datasets which are presented in detail in the remainder of this section.

\subsection{Experiment Design and Datasets}
\label{sec: experiment setup}

Consider two standard image classification datasets, \emph{MNIST} and \emph{CIFAR-10}. Let the FL system with $K = 10$ clients be designed to train convolutional neural networks (CNNs) over a total of $t = 50$ training rounds as discussed below. 
\begin{itemize}[leftmargin=*]
\setlength{\itemsep}{1.5ex}
\item \textbf{2-layer CNN on MNIST:} The first image classification dataset considered for the experiments is \emph{MNIST}, which contains 70,000 grayscale images of size $28 \times 28$ pixels containing handwritten digits $0$-$9$. Let the FL system be designed to train a simple CNN architecture comprising of two convolutional layers, followed by three fully connected layers. The MNIST dataset is split into 13 subsets, namely (i) 1000 random samples without repetition for $D_S$, (ii) 1000 random samples without repetition for $D_A$, (iii) 10,000 random samples without repetition for testing, and (iv) a maximum of 5800 random samples without repetition for $D_{k,t}$ at each client $k \in \mathcal{K}$.

\item \textbf{ResNet-18 on CIFAR-10:} The second image classification dataset used in our experiments is \emph{CIFAR-10}, which contains 60,000 color images of size $32 \times 32$ with 10 object classes. Let the FL system be designed to train a \emph{ResNet-18} \cite{resnet18}, a CNN architecture comprising of 18 convolutional layers with skip connections. The dataset is split into 13 subsets, namely (i) 1000 random samples without repetition for $D_S$, (ii) 1000 random samples without repetition for $D_A$, (iii) 10,000 random samples without repetition for testing, and (iv) a maximum of 4800 random samples without repetition for $D_{k,t}$ at each client $k \in \mathcal{K}$.
\end{itemize}

Furthermore, assume that each client performs local training over a total of two epochs, and choose the attacker's loss tradeoff factor $\lambda = 1$, loss tolerance $\epsilon = 0.01$, learning rates $\eta_s = \eta_a = 0.01$, and a belief update window of $T = 5$. A total of nine security vignettes were constructed by varying the fraction of compromised clients $\alpha$ to be one of the values in the set $\{0.1, 0.2, \cdots, 0.9\}$. Furthermore, let the CMP attacker's choice $a_{k,t} = a_t$ remains the same across all compromised clients $k \in \mathcal{K}_{comp}$. All experiments are initialized with $s_{k,0} = 1$ for all $k \in \mathcal{K}$ and $a_{k,0} = 0.5$, for all $k \in \mathcal{K}_{comp}$ respectively.

\subsection{State-of-the-Art Comparison and Evaluation Metrics}

The performance of \emph{StratFL} will be evaluated and compared against the following state-of-the-art methods: FedAvg \cite{fl}, Krum \cite{krum}, Trimmed Mean \cite{mean-median}, Coordinate-wise Median \cite{mean-median}, FLTrust \cite{fltrust} and FLANDERS \cite{flanders}, using the following evaluation metrics:

\begin{itemize}[leftmargin=*]
\setlength{\itemsep}{1.5ex}
\item \textbf{Accuracy Change:} The extent of damage caused by the CMP attacker at the global model can be quantified by \emph{accuracy change}, 
\begin{equation}
\Delta\Psi = \Psi_{t}(\emptyset) - \Psi_{t}(\mathcal{K}_{comp}),
\end{equation}
where $\Psi_{t}(\emptyset)$ is the accuracy obtained by the global model at round $t$ in the absence of a CMP attacker (i.e. when $\mathcal{K}_{comp} = \emptyset$ is an empty set.), and $\Psi_{t}(\mathcal{K}_{comp})$ is the accuracy obtained by the global model at round $t$ in the presence of a CMP attacker who controls clients in $\mathcal{K}_{comp}$. 

\item \textbf{Mean-squared belief deviation:} The success of belief updating algorithms at both the server and the attacker can be quantified based on the average of the squared deviation of $(1 - s_{k,t})$ from the client type $\gamma_k$ as
\begin{equation*}
MSBD = \frac{1}{T} \sum_{t=1}^{T} \left\{ \frac{1}{K} \sum_{k=1}^{K} \big( 1 - s_{k,t} - \gamma_{k}\big)^{2} \right\}.
\end{equation*}
\label{Eqn: SBD}
\end{itemize}

\section{Results and Discussion}

\begin{figure}[!t]
\centering
\begin{subfigure}{0.49\textwidth}
\centering
\includegraphics[width=\textwidth]{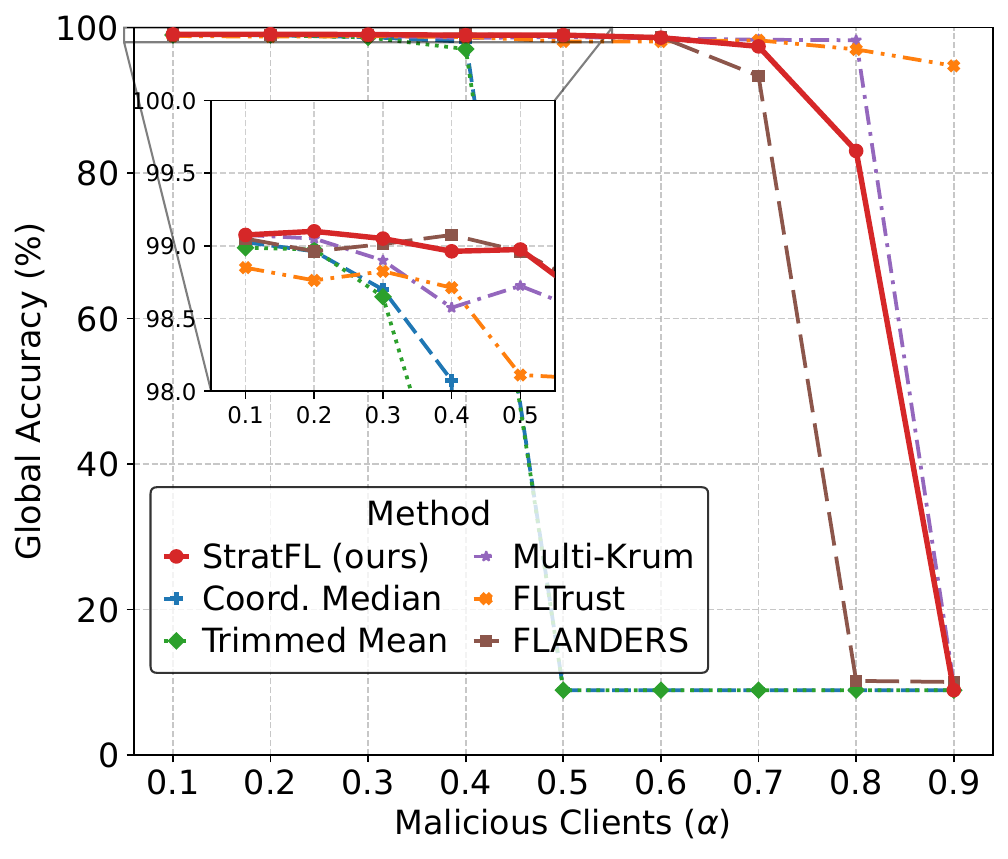}
\label{fig:1a}
\end{subfigure}
\begin{subfigure}{0.49\textwidth}
\centering
\includegraphics[width=\textwidth]{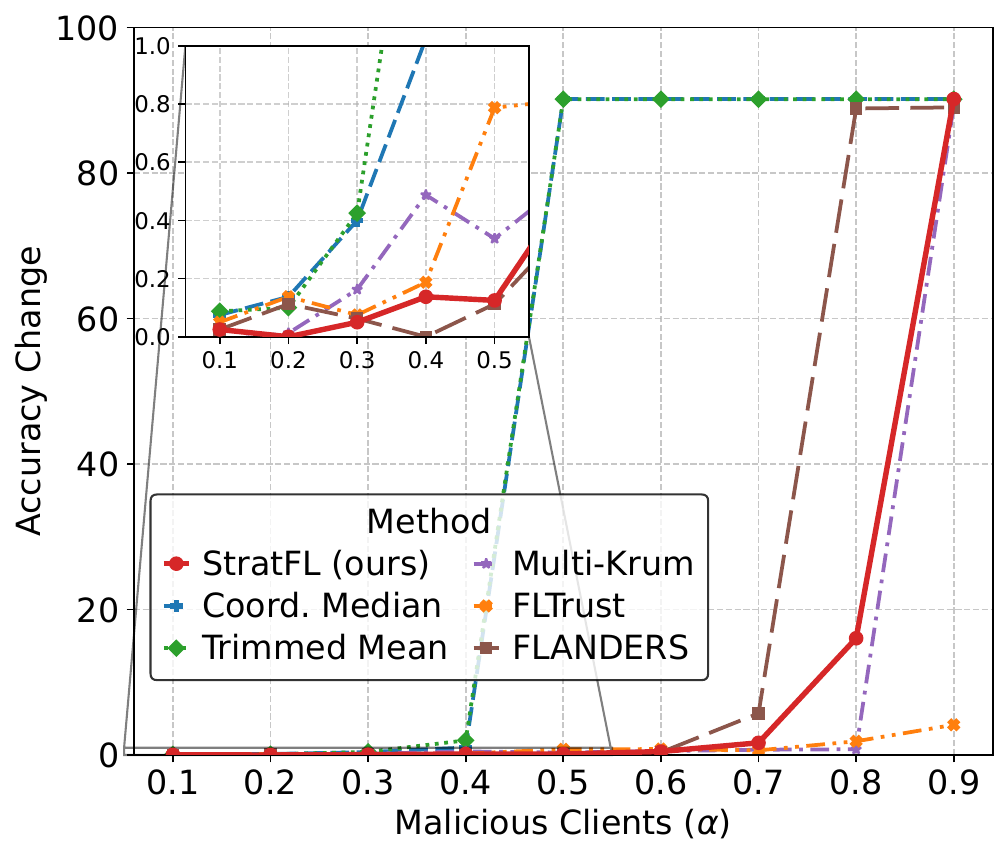}
\label{fig:1b}
\end{subfigure}
\caption{Comparison of Global Model Accuracy after 50 Training Rounds on MNIST Dataset}
\label{mnist-accuracy}
\end{figure}

\begin{figure}[!t]
\centering
\begin{subfigure}{0.49\textwidth}
\centering
\includegraphics[width=\textwidth]{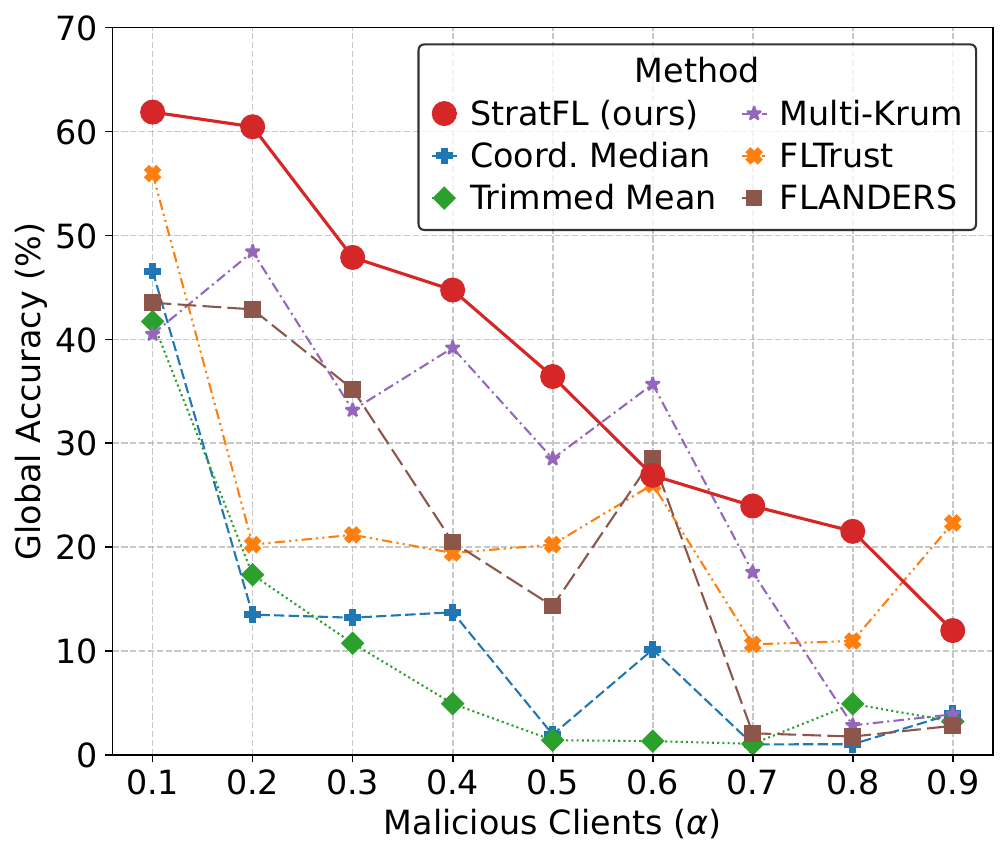}
\label{fig:2a}
\end{subfigure}
\begin{subfigure}{0.49\textwidth}
\centering
\includegraphics[width=\textwidth]{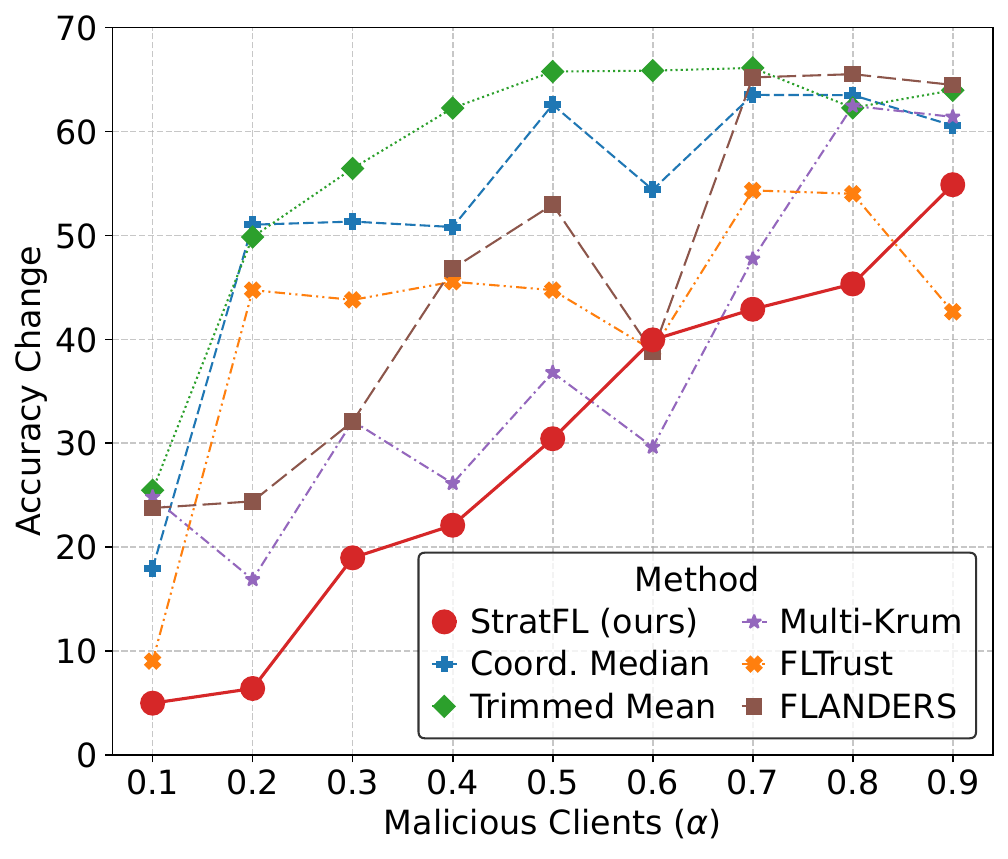}
\label{fig:2b}
\end{subfigure}
\caption{Comparison of Global Model Accuracy after 50 Training Rounds on CIFAR-10 Dataset}
\label{cifar10-accuracy}
\end{figure}

\begin{figure}[!t]
\centering
\includegraphics[width=\textwidth]{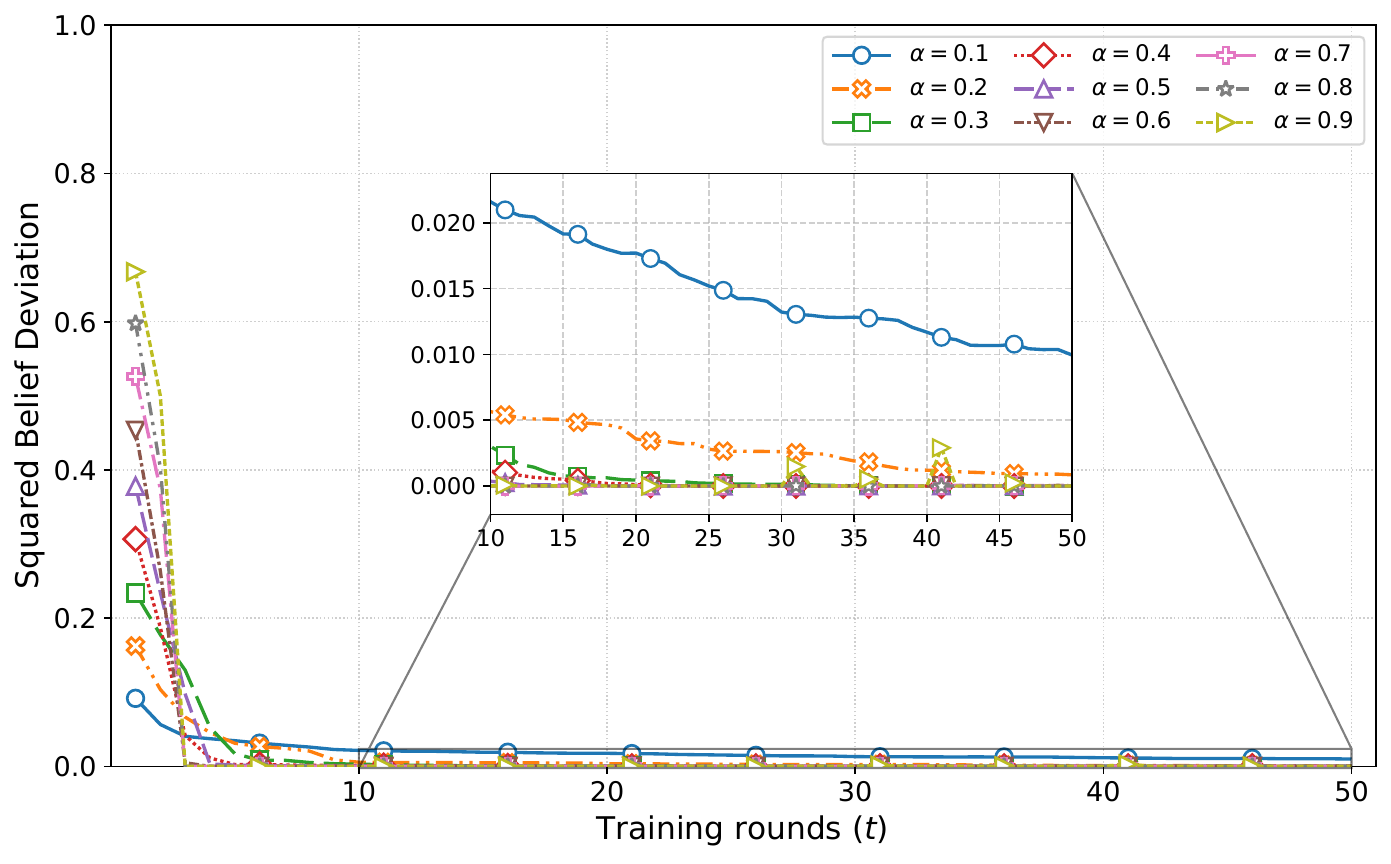}
\caption{Empirical Convergence of Mean Squared Belief Deviation on CIFAR-10 Dataset across Different Security Vignettes}
\label{mnist-sdb}
\end{figure}

\begin{figure}[!t]
\centering
\includegraphics[width=\textwidth]{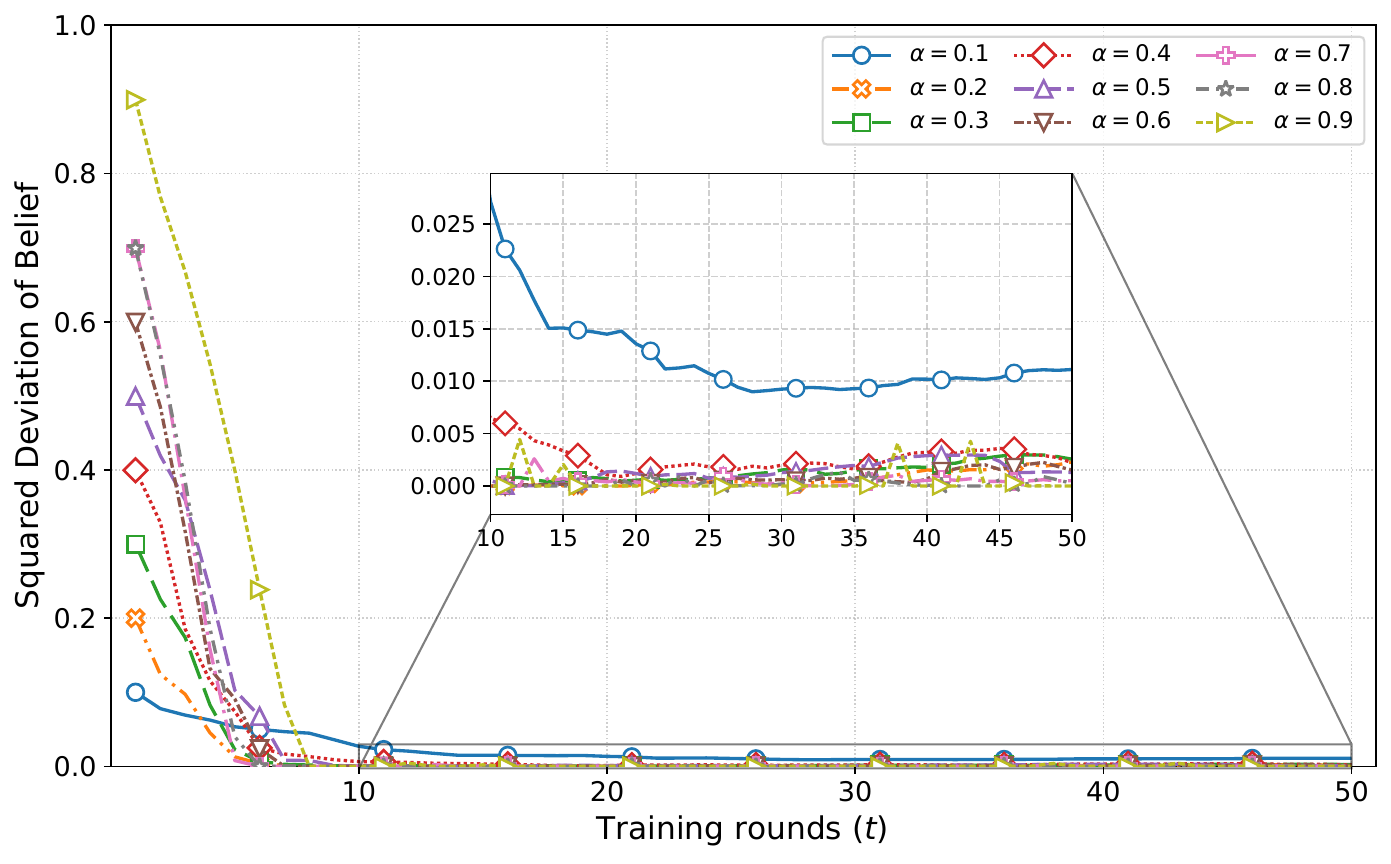}
\caption{Empirical Convergence of Mean Squared Belief Deviation on CIFAR-10 Dataset across across Different Security Vignettes}
\label{cifar10-sdb}
\end{figure}

\subsection{Accuracy Change across Different Security Vignettes}

Figures \ref{cifar10-accuracy} and \ref{mnist-accuracy} illustrate the global model accuracy and the corresponding accuracy change due to \emph{StratFL} and other state-of-the-art methods in MNIST and CIFAR-10 datasets respectively. \emph{StratFL} clearly outperforms all the baseline methods for any $\alpha \leq 0.7$. However, if the majority of the network is compromised by the CMP attacker, then the only method that survives is FLTrust. FLTrust outperforms \emph{StratFL} for large values of $\alpha$ because of its server-side training, which is an unreasonable assumption in most FL deployments.  

Specifically, Figure \ref{mnist-accuracy} presents the accuracy and accuracy change on MNIST dataset. \emph{StratFL}'s accuracy tracks the baselines until only $1$ honest client remains. The inset of Figure \ref{mnist-accuracy} presents that all methods stay within the range of $\approx 99\%$ accuracy, through $\alpha =  0.4$. Coordinate-Median and Trimmed-Mean fall to $10\%$ at $\alpha = 0.5$, which coincides with the breaking point of these methods: once the adversary controls half of the coordinates, the per-coordinate statistic follows the adversary. \emph{StratFL} stays above $99\%$ accuracy through $\alpha = 0.7$, declining only at $\alpha = 0.8$ and reaches $10\%$ at $\alpha = 0.9$. With $10$ clients, $\alpha = 0.9$ leaves only $1$ honest client leaving little to no gradient information for the algorithm to learn. 

On the other hand, Figure \ref{cifar10-accuracy} presents the results on accuracy and accuracy change for the CIFAR-10 dataset. Here, the accuracy of every method decreases as $\alpha$ increases. However, \emph{StratFL} still leads across all vignettes of $\alpha$. Trimmed-Mean reaches $0\%$ accuracy by $\alpha = 0.5$; FLTrust and Multi-Krum also decline ahead of \emph{StratFL} by $\alpha = 0.5$. This trend holds across on CIFAR-10 because the gradient update  moves off the descent direction of the honest clients. The right panel in Figure \ref{cifar10-accuracy} also reports the same effect as accuracy change is the lowest among all the methods through $\alpha = 0.5$.

Under the presence of a CMP attacker, the defense mechanisms fail due to two reasons: (i) admitting erroneous updates into the aggregation function, or (ii) removal of every erroneous update and retaining too few honest updates to  estimate a gradient descent direction. The accuracy degradation (as $\alpha$ increases) is therefore due to the lack of honest gradient information at the server's side, and is independent of \emph{StratFL}'s belief updating mechanism.

\subsection{Convergence of Belief Updating}

Figures \ref{mnist-sdb} and \ref{cifar10-sdb} present the Mean Squared Belief Deviation ($MSBD$) on MNIST and CIFAR-10 datasets, across $\alpha = \{0.1, \cdots, 0.9\}$. It can be observed that \emph{StratFL} converges withing 10 rounds across all $\alpha$ on both datasets. After convergence, this trajectory holds without any deviation. The observations show that the belief update mechanism converges against all the tested values of $\alpha$. Once converged, the CMP attacker's $\boldsymbol{a}_{t}$ is no longer able to influence the server's choice $\boldsymbol{s}_{t}$. 

Convergence follows from the structure of the updates. The server accumulates belief $\pi_{t}$ over the training rounds and maps the loss inequality indicator $c_{k,t}$ to its updates $s_{k,t}$ (Section \ref{sec:server-beliefupdate}). At the beginning of training, $s_{k,t}$ values are at $1$, resulting in higher $MSBD$ values. This is consistent with the assumption that the server starts without any prior knowledge that distinguishes client types. As training progresses, the accumulation of $c_{k,t}$ values enables the separation of clients. Specifically, the clients are separated by those which increase the server's validation loss $D_{S}$ beyond the $\epsilon$ margin and those which decrease it. This separation results in $s_{k,t}$ values moving towards convergence thereby reducing the $MSBD$. The training round at which the $s_{k,t}$ values stop changing marks the round at which the server's converges regarding its choice updates. 

The configuration at $\alpha = 0.1$ retains an $MSBD$ value that exceeds every other configuration on the both datasets ($0.010$ on MNIST and $0.011$ on CIFAR-10). Specifically, at $\alpha = 0.1$, there is one malicious client placed among $10$ clients. Therefore, the deviation here comes from $1$ single client at every training round $t$ and it persists. As $\alpha$ increases, the deviation that the server observes also increases thereby leading the $MSBD$ to reach $0$ at the end of training ($t=50$). 

The trajectory holds even after convergence. Under $\lambda$, the adversary trades the attack impact against covertness. An increase in $a_{k,t}$ raises $c_{k,t}$ at the server thereby driving the server to decrease $s_{k,t}$ while a drop in $a_{k,t}$ lowers the damage that the adversary intends to make. In either scenarios, the $MSBD$ value does not lower once $s_{k,t}$ converges. Therefore, the adversary holds $a_{k,t}$ and the CMP attack fails to impact the server's model aggregation.  

Across all $\alpha$ values, $MSBD$ stays under $0.03$ by $t = 10$ and $0.011$ by $t=50$. Empirically, \emph{StratFL} converges by $t = 5$ on MNIST and $t = 10$ on CIFAR-10, across all vignettes of $\alpha$ 

\section{Conclusion and Future Work}

In summary, an adversarial search algorithm \emph{StratFL} was proposed to counter covert model poisoning threats in federated learning. Experiments were designed to evaluate it against state-of-the-art methods. We found that \emph{StratFL} exhibits 10\% improvement over baseline methods in terms of accuracy while also preserving accuracy under and at most 70\% malicious clients on both of the MNIST and CIFAR-10 datasets (at which the baseline methods collapse). In addition, \emph{StratFL}'s beliefs converge within the first 10 rounds of training, a dimension of robustness that other defense mechanisms do not capture. 

In the future, we will explore other APTs such as data poisoning, backdoor attacks against federated learning networks.

%
%
%
\bibliographystyle{splncs04}
\bibliography{references}
%




\end{document}